\documentclass[aps,prd,preprint,
superscriptaddress]{revtex4}

\usepackage{graphicx}
\usepackage{amsmath}
\usepackage{bm}
\usepackage{color}
\usepackage{setspace}
\usepackage{comment}

\usepackage{soul}

\def\ee{\end{eqnarray}}

\def\=:{=\hspace{-.7em}\raisebox{1.1ex}{.}\hspace{.1em}\raisebox{-0.2ex}{.} }

\def\ee{\end{eqnarray}}

\def\=:{=\hspace{-.7em}\raisebox{1.1ex}{.}\hspace{.1em}\raisebox{-0.2ex}{.} }

\newcommand {\beq}{\begin{eqnarray}}
\newcommand {\eeq}{\end{eqnarray}}

\newcommand {\1}[1]{\frac{1}{#1}}

\newcommand {\del}{\partial}

\newcommand{\vs}[1]{\vspace{#1 mm}}

\begin{document}


\title{Toroidal domain walls as Hopfions
}


\author{Michikazu Kobayashi}
\affiliation{
Department of Basic Science, University of Tokyo, Komaba 3-8-1, Meguro-ku, Tokyo 153-8902, Japan}
\affiliation{
Department of Physics, Kyoto University, Oiwake-cho, Kitashirakawa, Sakyo-ku, Kyoto 606-8502, Japan}
\author{Muneto Nitta}
\affiliation{Department of Physics, and Research and Education Center for Natural 
Sciences, Keio University, Hiyoshi 4-1-1, Yokohama, Kanagawa 223-8521, Japan
}



\date{\today}
\begin{abstract}
We construct stable domain walls in a shape of a torus 
in the Faddeev-Skyrme model with a quadratic potential term admitting two discrete vacua.  
The phase modulus of the domain wall is twisted 
$P$ and $Q$ times  
along the toroidal and poloidal cycles of the torus, 
respectively,
where $Q$ represents the constituent lump charge.
We find that these solutions carry the Hopf charge 
$PQ$
and can be regarded as Hopfions.
The toroidal domain walls are 
characterized by the two topological charges 
$(P,Q)$, 
unlike the conventional Hopfions characterized by 
only the Hopf charge.
We explicitly construct stable solutions of 
the toroidal domain walls 
with $1 \leq P \leq 3$, and $1 \leq Q \leq 3$, 
yielding the Hopf charges $1,2,3,4,6,9$.

\end{abstract}

\maketitle

\section{Introduction}

Topological solitons are ubiquitous in various subjects in physics from condensed matter physics \cite{Volovik2003} to high energy physics \cite{Manton:2004tk} and cosmology \cite{Vilenkin:2000}. One of the most simplest and most frequently appearing solitons are kinks in $d=1+1$ dimensions 
or domain walls $d=2+1$ or larger dimensions, 
which are present in various 
condensed matter systems such as
optical fibers, ferromagnets, charge density waves,  Bose-Einstein condensates, helium superfluids and so on.  
Domain walls are also studied in high energy physics 
and cosmology. 
In supersymmetric theories,
domain walls are Bogomol'nyi-Prasad-Sommerfield states
preserving a half of supersymmetries 
and are stable against perturbative and non-perturbative quantum corrections. 
Domain walls have also been studied in the context of brane-world scenario.

In $d=3+1$ dimensions, domain walls have $2+1$ dimensional world-volume.
Flat and linear world-volume are usually the most stable ground state of the domain walls. A question arises. 
Is it always the case that domain walls have flat shapes? 
Can they be spherical or of Riemann surfaces? 

In this paper, we construct stable domain walls of 
a torus shape. 
The model which we consider is an $O(3)$ sigma model 
with a four derivative (Skyrme) term, 
known as the Faddeev-Skyrme (FS) model 
\cite{Faddeev:1975,Faddeev:1996zj}, 
with 
the ferromagnetic or Ising-type potential, that is,  
a potential term admitting two discrete vacua
\cite{Nitta:2012kk}. 
The FS model without a potential term 
is known to admit Hopfions, {\it i.e.} solitons 
with the Hopf charge $\pi_3(S^2) \simeq {\bf Z}$  
\cite{Gladikowski:1996mb,Faddeev:1996zj,Battye:1998pe,
Hietarinta:2000ci,Sutcliffe:2007ui,Radu:2008pp}.
In particular, Hopfions with the Hopf charge 7 or 
higher were found to have knot structures  
\cite{Battye:1998pe,Hietarinta:2000ci,Sutcliffe:2007ui}.
Hopfions in the FS model with different 
potential terms were also studied before  
\cite{Foster:2010zb,Harland:2013uk,Battye:2013xf}, 
but the ferromagnetic potential was not studied. 
From the construction, our toroidal domain walls carry the Hopf charge 
\cite{deVega:1977rk,Gladikowski:1996mb} 
and consequently they can be regarded as Hopfions.

The $O(3)$ model with a potential term with two discrete vacua 
admits a domain wall solution 
interpolating the two discrete vacua  \cite{Abraham:1992vb,Arai:2002xa}.
The domain wall has the position and $U(1)$ phase moduli. 
In $d=2+1$
a closed domain line or a wall ring 
with the $U(1)$ modulus twisted $Q$ times  
is nothing but a lump 
with the lump charge $Q$ \cite{Polyakov:1975yp} 
in the absence of  the Skyrme term \cite{Nitta:2012kj,Nitta:2012kk}. 
While this wall ring is unstable to shrink 
in the absence of  the Skyrme term, 
it is stable in the presence of the Skyrme term.
This is nothing but a baby Skyrmion \cite{Piette:1994ug,Weidig:1998ii}.
In $d=3+1$, the twisted wall ring is linearly extended along one direction 
to a stable tube.  
In this paper, we further make a closed loop of the tube with 
the $U(1)$ modulus twisted $P$ times, 
in order to construct 
a stable toroidal domain wall. 
Therefore
the $U(1)$ modulus is twisted $P$ and $Q$ times 
along the toroidal and poloidal cycles of the torus, 
respectively. 
This solution carries the Hopf charge $C=PQ$ \cite{deVega:1977rk,Gladikowski:1996mb,Hietarinta:2000ci},
and is also topologically equivalent to a Q-torus studied in \cite{Bolognesi:2007zz} 
in which time-dependent stationary solutions 
in the model without the Skyrme term were discussed.
We explicitly construct stable solutions of 
the toroidal domain walls 
with $1 \leq P \leq 3$, and $1 \leq Q \leq 3$, 
yielding the Hopf charges $1,2,3,4,6,9$.
We conclude that 
the Hopfions with the Hopf charge $C$ in our model are
further classified 
into a set of infinite series characterized by 
the two topological charges $(P,Q)$ with $C=PQ$, 
where $Q$ is the constituent lump charge.

This is in contrast to the conventional Hopfions 
as knot solitons classified only by the only Hopf charge 
$C$ in the original FS model, 
in which configurations with different 
sets $(P,Q)$ with $C=PQ$ 
are topologically equivalent and can be deformed 
to each other \cite{Hietarinta:2000ci}. 
In our model, configurations with different 
sets $(P,Q)$ with $C=PQ$ 
are  topologically distinct and are all stable (or metastable)
at absolute (or local) minima of the energy, 
because of the existence of a potential barrier among them.

This paper is organized as follows. 
 In Sec.~\ref{sec:model}, 
after our model is given, 
we construct twisted domain wall rings in $d=2+1$ dimensions. 
They can be linearly extended to domain wall tubes in $d=3+1$.
In Sec.~\ref{sec:toroidal-wall}, we construct 
twisted closed domain wall tubes, 
resulting in toroidal domain walls. 
The $U(1)$ modulus is twisted along the 
both cycles of the torus.  
Section \ref{sec:summary} is devoted to 
discussions. 

\section{The model and solitons in $1+1$ and $2+1$ dimensions\label{sec:model}}
\bigskip
Let ${\bf n} (x)= (n_1(x),n_2(x),n_3(x))$  
be a unit three vector of scalar fields 
with a constraint ${\bf n} \cdot {\bf n} = 1$.
The Lagrangian of our model is given by
($\mu=0,1,2,3$)
\beq
&& {\cal L} = \1{2} \del_{\mu}{\bf n}\cdot \del^{\mu} {\bf n} 
 - {\cal L}_4({\bf n})
 - V({\bf n}). \quad  
 \label{eq:Lagrangian}
\eeq
Here, the four-derivative (Faddeev-Skyrme) term  
is given by
\beq
{\cal L}_4 ({\bf n})
= \kappa F_{\mu\nu}^2
= \kappa  \left[{\bf n} \cdot 
 (\partial_{\mu} {\bf n} \times \partial_{\nu} {\bf n} )\right]^2
= \kappa (\partial_{\mu} {\bf n} \times \partial_{\nu} {\bf n} )^2  
\eeq
with a ``field strength," 
\beq
F_{\mu\nu}={\bf n}\cdot 
(\partial_{\mu} {\bf n} \times \partial_{\nu} {\bf n} ). 
\eeq 
We take the potential term \cite{Abraham:1992vb,Weidig:1998ii}
\beq
V({\bf n}) = m^2(1-n_3^2),  \label{eq:pot}
\eeq 
admitting two discrete vacua $n_3=\pm 1$ 
in order to allow a domain wall solution. 
In condensed matter physics, 
this potential is well known in ferromagnets, 
so we call it ferromagnetic of the Ising-type.
The energy density of static configurations is  
\beq
\mathcal{E} =
\1{2} ( \del_a{\bf n}\cdot \del^a {\bf n} ) 
+ {\cal L}_4({\bf n}) + V({\bf n})
\eeq
with $a=1,2,3$.


In $d=1+1$ dimensions, 
a kink (an anti-kink)  solution interpolating these two vacua in the absence of the Skyrme term 
is given by \cite{Abraham:1992vb,Arai:2002xa,Nitta:2012wi}
\beq
\label{eq:wall} 
&& \theta (x^1) = 2 \arctan \exp (\pm \sqrt 2 m (x^1 -X)),  
  \quad 0 \leq \theta \leq \pi  , \nonumber \\
&&  n_1 = \cos \alpha \sin \theta (x^1) , \quad 
      n_2 = \sin \alpha  \sin \theta (x^1) , \quad
      n_3 = \cos \theta (x^1),
\eeq
with a phase modulus 
$\alpha$ ($0 \leq \alpha < 2\pi$) and 
the translational modulus $X \in {\bf R}$ 
of the domain wall. 
A domain wall solution in the presence of the Skyrme term was studied in Ref.~\cite{Kudryavtsev:1997nw}.
In $d=2+1$ and $d=3+1$ dimensions, the kink can be linearly extended to a domain line and a domain wall, respectively, with a world-volume.

\begin{figure}
\begin{center}
\vspace{-11cm}
\includegraphics[width=1.0\linewidth,keepaspectratio]{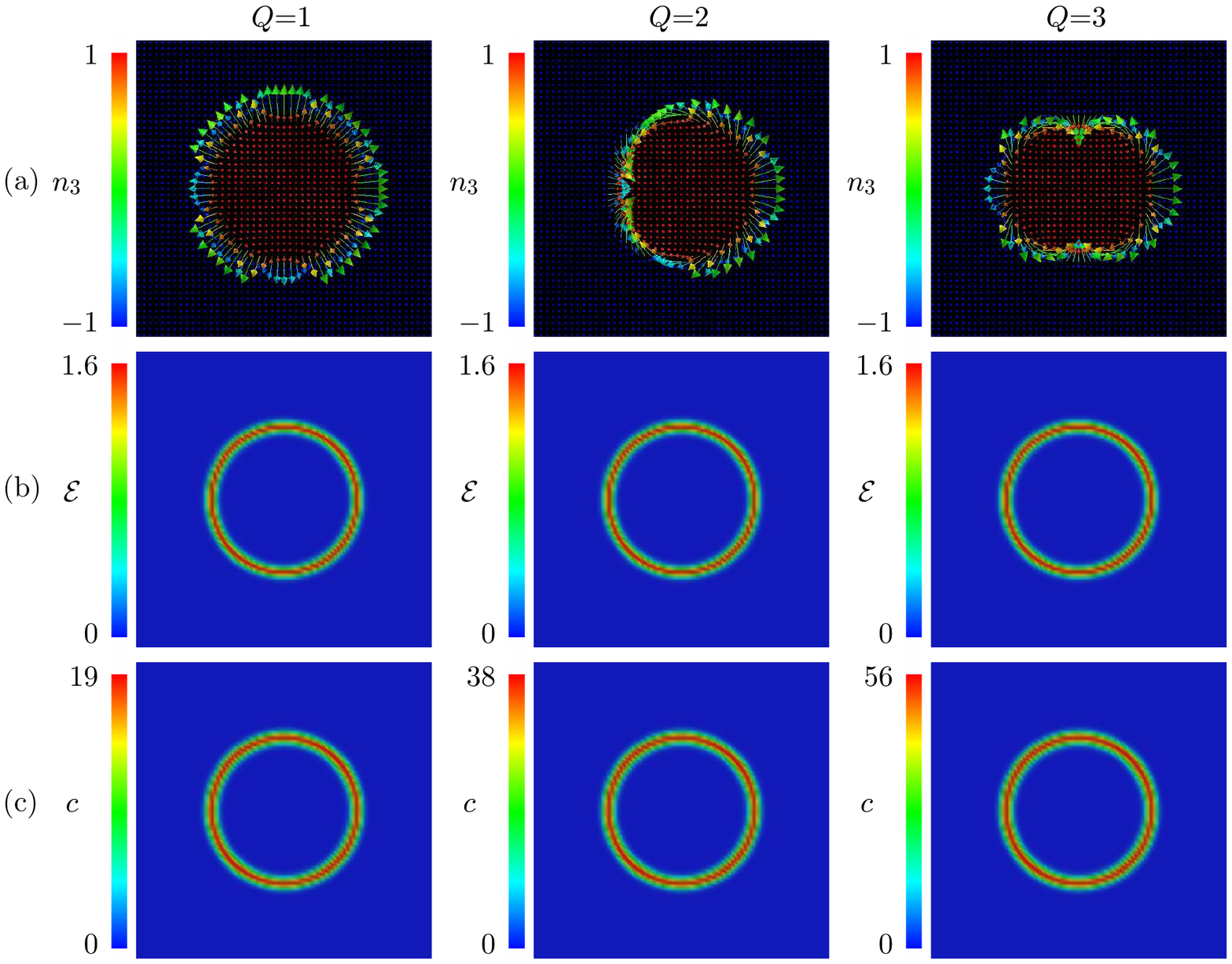}
\caption{A twisted domain wall ring as a baby Skyrmion with the lump charges of $Q = 1$ (left), $Q = 2$ (middle), and $Q = 3$ (right).
(a): The texture ${\bf n}(x)$. Color of each arrow show the value of $n_3$.
(b): The total energy density $\mathcal{E}$. 
(c): The topological lump charge density $c \equiv  F_{12} / (4 \pi)$.
The numerical box satisfies the Neumann boundary condition.
As numerical parameters, we fix $m^2 = 8000$ for all figures, and $\kappa / L^4 = 0.4 \times 10^{-3}$ for left figures $(Q = 1)$, $\kappa / L^4 = 0.1 \times 10^{-3}$ for middle figures $(Q = 2)$, and $\kappa / L^4 = 0.5 \times 10^{-4}$ for right figures $(Q = 3)$ where $L$ is the size of the numerical box.
\label{fig:wall-ring}}
\end{center}
\end{figure}
In $d= 2+1$ dimensions, 
one can consider a closed domain line. 
If the $U(1)$ modulus $\alpha$ of the domain line 
winds $Q$ times along the wall ring,  
it carries the topological lump charge of
 $\pi_2(S^2) \simeq {\bf Z}$ \cite{Nitta:2012kj}, 
given by
\beq
 Q
= \1{4 \pi}  \int d^2x\: F_{12} 
= \1{4 \pi}  \int d^2x\: {\bf n}\cdot 
(\partial_1 {\bf n} \times \partial_2 {\bf n} )
=
 \1{4 \pi}  \int d^2x\:
\epsilon_{ijk}
n_i \partial_{1} n_j  \partial_{2} n_k . \label{eq:lump-charge}
\eeq 
This twisted closed wall line is prevented 
against shrinking by the presence of the Skyrme term  
\cite{Kobayashi:2013ju}.  
It is nothing but a baby Skyrmion \cite{Piette:1994ug}.
In Fig.~\ref{fig:wall-ring}, 
we give our numerical solutions of 
twisted domain wall rings with the lump charges 
$Q= 1$, $Q = 2$, and $Q = 3$ constructed by a relaxation method. 
One can see that the topological lump charge density as well as the energy density is uniformly distributed along the rings.
The multi-winding solutions with $Q=2$ and $Q=3$ are stable 
and are not split into $Q=1$ rings.
The size of the lump soliton becomes larger for the higher lump charge $Q$ within the fixed $\kappa$ and $m$.
In our numerical simulation, to make the sizes of lump solitons within different $Q$ almost same, we use the smaller $\kappa$ for higher $Q$. 

\section{Toroidal domain walls in $3+1$ dimensions
\label{sec:toroidal-wall}}

\begin{figure}
\begin{center}
\vspace{-10cm}
\includegraphics[width=0.6\linewidth,keepaspectratio]{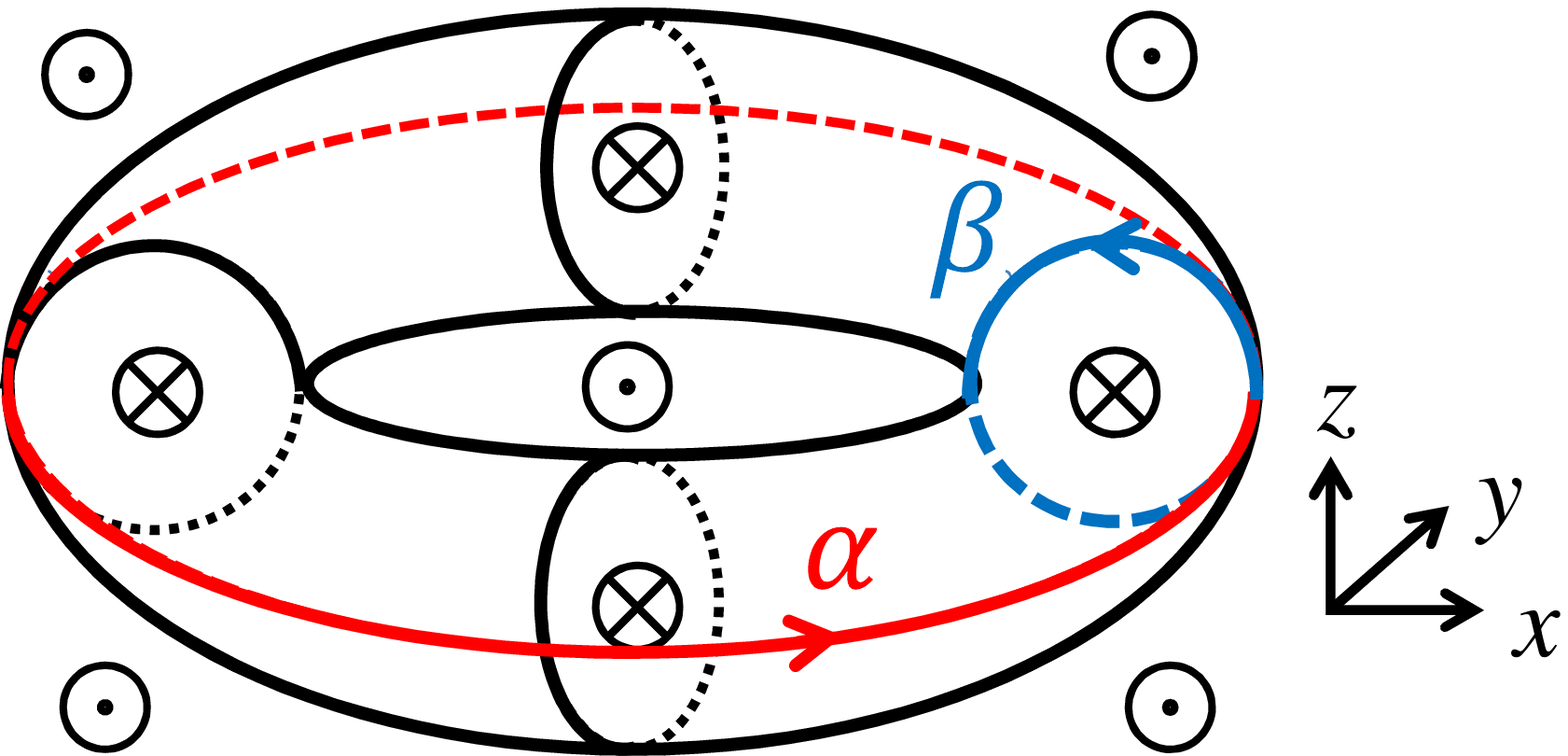}
\end{center}
\caption{The two cycles of the torus. 
The toroidal and poloidal cycles are 
denoted by $\alpha$ and $\beta$, respectively. 
The $\odot$ and $\otimes$ denote the vacua at the north and south 
of the target space $S^2$, respectively.  
The $U(1)$ modulus is twisted $P$ and $Q$ times 
along the cycles $\alpha$ and $\beta$, respectively. \label{fig:cycles}}
\end{figure}

The domain wall ring can be linearly extended to a domain wall tube in $d=3+1$ dimensions. 
Here we further consider a closed domain wall tube, 
{\it i.e.}, a toroidal domain wall. 
Such a toroidal domain wall is unstable unless 
the $U(1)$ modulus is twisted along the torus. 
Therefore, we twist the $U(1)$ modulus along the torus, 
resulting in a doubly twisted toroidal domain wall.
First, let us put a charge $Q$ (or $-Q$) lump in 
the $y>0$ (or $y<0$) region of the $y$-$z$ plane, 
\beq
 Q = \1{4\pi} \int_{x=0,y>0} dy dz\: F_{23} 
    = - \1{4\pi} \int_{x=0,y<0} dy dz\: F_{23}.
\eeq
Then,  let us rotate it around the $z$-axis with twisting 
the $U(1)$ modulus $P$ times in the $2\pi$ rotation. 
Here, $P$ can be expressed as 
\beq
 2P = \1{4\pi} \int_{z=0} dx dy\: F_{12},  
\eeq
where the factor $2$ in the left hand side is because of 
the two rings of the cross section of a torus and the $x$-$y$ plane.
In the final configuration, 
the $U(1)$ modulus is twisted $P$ and $Q$ times along 
the $\alpha$ and $\beta$ cycles of the torus, respectively, as in Fig.~\ref{fig:cycles}. 

Below, we show that 
a toroidal domain wall characterized by the two integers 
$(P,Q)$ carries the Hopf charge 
of $\pi_3(S^2)\simeq {\bf Z}$ defined by 
\beq
 C = \1{4 \pi^2}  \int d^3x\: \epsilon^{\mu\nu\rho} 
F_{\mu\nu} A_{\rho} ,
\eeq
with a ``gauge field" $A_{\mu}$ satisfying $\del_{\mu}A_{\nu}-\del_{\nu}A_{\mu}=F_{\mu\nu}$ 
\cite{deVega:1977rk}, 
and that 
the Hopf charge in fact coincides with the product of 
the numbers of the twisting of the $U(1)$ phase along 
the $\alpha$ and $\beta$ cycles:
\beq
 C = PQ.
\eeq
\begin{figure}
\centering
\vs{-5cm}
\includegraphics[width=0.7\linewidth]{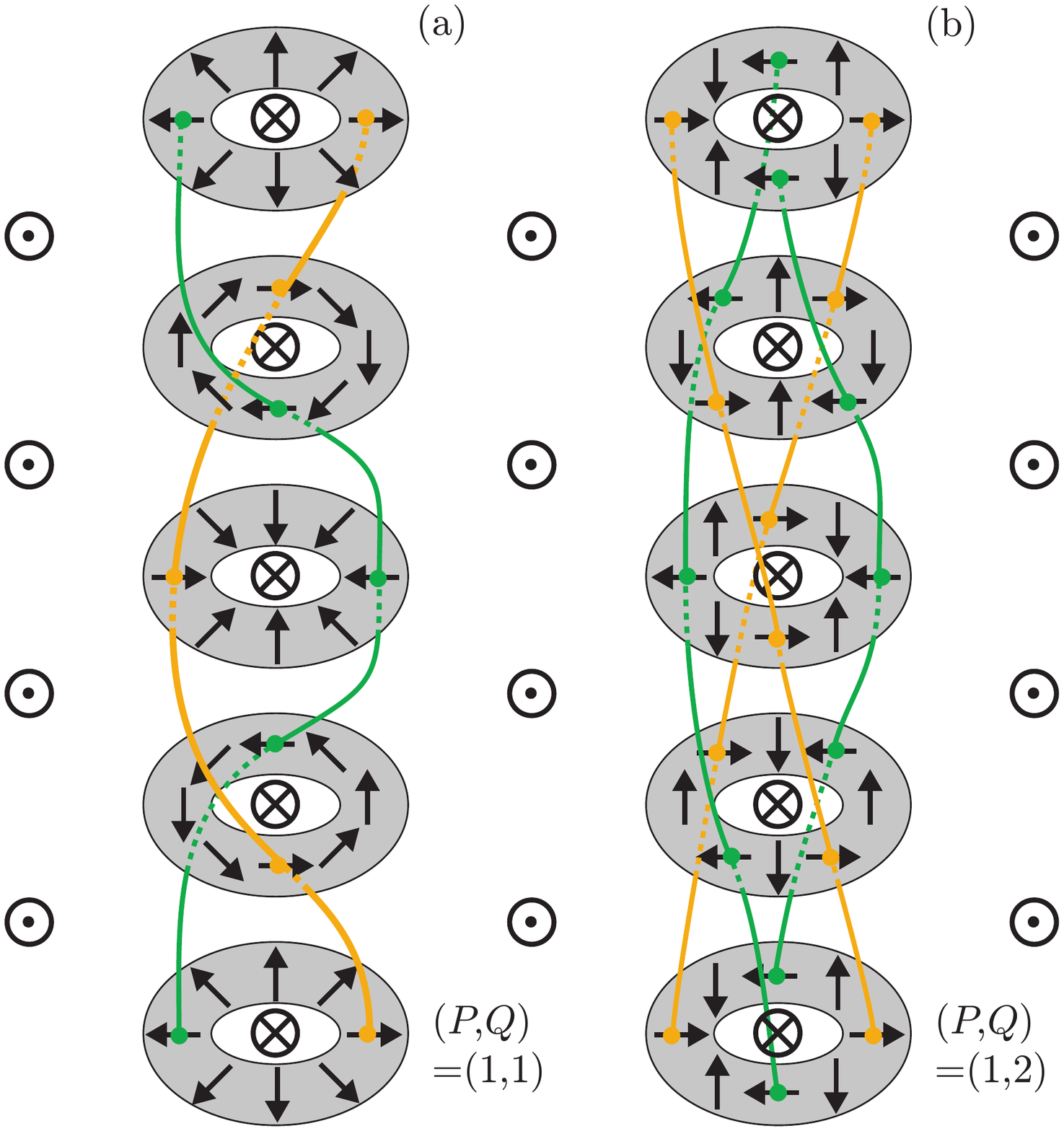}
\caption{\label{fig:cylinder} A cut along the $\beta$ cycle 
of the toroidal domain walls.
The gray $n_3 = 0$ surfaces represent domain wall tubes 
that separate the two vacua  
${\bf n}=(0,0,1)$ denoted by $\odot$ outside the tube 
and ${\bf n}=(0,0,-1)$ denoted by $\otimes$ 
inside the tube. 
Along the tubes, there are sequences of lumps with the lump charges of (a) $Q = 1$ and  (b) $Q = 2$. From the bottom to the top, ${\bf n}$ rotates by $2 \pi P$ ($P = 1$) in the $n_1$--$n_2$ plane, defining the number of twists $P$.
The yellow and green curves indicate the preimages of 
${\bf n}= (1,0,0)$ and ${\bf n}= (-1,0,0)$ along the tubes respectively.
We find that the preimages of ${\bf n} = (1,0,0)$ and ${\bf n} = (-1,0,0)$ are linked once in (a) ($Q=1$) and twice in (b) ($Q=2$).
}
\end{figure}

To show this, we promote configurations with 
the target space $S^2$ to 
those with $S^3$ by the Hopf map.
With introducing two complex scalar fields 
$\phi^T=(\phi^1,\phi^2)$ satisfying 
$|\phi^1|^2+|\phi^2|^2=1$, 
the three-vector scalar fields $n_i$ 
can be written by the Hopf map 
\beq
  n_i = \phi^\dagger \sigma_i \phi , \quad (i=1,2,3)
\label{eq:Hopf-map}
\eeq
by using the Pauli matrices $\sigma_i$.
Note that $\phi$ parametrize $S^3 \simeq SU(2)$ because of this constraint.
As an initial configuration, we consider an ansatz
\begin{align}
\begin{array}{c}
\displaystyle \phi^T = \begin{pmatrix} \sin\left[ \cos^{-1}\{\sin f(r) \sin\theta\} \right] e^{- i Q \tan^{-1} \frac{\sin f(r) \cos\theta}{- \cos f(r)}} , & \cos\left[ \cos^{-1}\{\sin f(r) \sin\theta\} \right] e^{i P \phi} \end{pmatrix} , \vspace{5pt} \\
\displaystyle 0 \leq r < \infty,\quad
0 \leq \theta \leq \pi, \quad
0 \leq \phi \leq 2 \pi,
\end{array}
\label{eq:ansatz}
\end{align}
where a monotonically decreasing function $f(r)$ satisfies
the boundary condition 
\begin{align}
f(r \to 0) \to \pi, \quad
f(r \to \infty) \to 0.
\end{align}
From the Hopf map in Eq.~(\ref{eq:Hopf-map}), 
we have
\begin{align}
\begin{array}{c}
\displaystyle
n_1 = \sin\Theta \cos(Q \Phi), \quad
n_2 = \sin\Theta \sin(Q \Phi), \quad
n_3 = \cos\Theta \vspace{5pt}, \\
\displaystyle
\Theta = \tan^{-1} \frac{\sqrt{X^2 + Y^2}}{Z}, \quad
\Phi = \tan^{-1} \frac{Y}{X},
\end{array} \label{eq:Hopfion}
\end{align}
with $X$, $Y$, and $Z$ defined by
\begin{align}
\begin{split}
X &= \sin\theta\left\{ \sin f(r) \sin(P \phi / Q) + \{ 1 - \cos f(r) \} \cos\theta \cos(P \phi / Q) \right\}, \\
Y &= \sin\theta\left\{ - \sin f(r) \cos(P \phi / Q) + \{ 1 - \cos f(r) \} \cos\theta \sin(P \phi / Q) \right\}, \\
Z &= \cos^2\theta + \cos f(r) \sin^2\theta.
\end{split}
\end{align}
The configuration given in Eq.~\eqref{eq:Hopfion} 
is isomorphic to a toroidal domain wall with $(P, Q)$ and its Hopf charge $C$ can be obtained 
through the Hopf map in Eq.~(\ref{eq:Hopf-map})
from the Skyrme charge $\pi_3(S^3) \simeq {\bf Z}$ 
of the fields $\phi$ in Eq.~\eqref{eq:ansatz}:
\begin{align}
\begin{split}
C &:= \frac{1}{2 \pi^2} \int d^3x\: \epsilon^{ABCD} 
m^A \partial_1 m^B \partial_2 m^C \partial_3 m^D \\
&= \frac{1}{2 \pi^2} \int_0^{\infty} dr\: \int_0^\pi d\theta\: \int_0^{2 \pi} d\phi\: r^2 \sin\theta \frac{P Q f^\prime(r) \sin^2 f(r)}{r^2} \\
&= \frac{P Q}{\pi} \int_0^\infty dr\: \frac{d}{dr} \left\{ f(r) - \sin f(r) \cos f(r) \right\} \\
&= P Q,
\end{split}
\end{align}
with $\phi^1 = m^1 + i m^2$ and $\phi^2 = m^3 + i m^4$ 
($A,B,C,D=1,2,3,4$).

We further note that a preimage of a point in the target space 
$S^2$ is a closed loop in the real space, 
and that the Hopf charge is equivalent to the linking number 
of two preimages of two points in the target space \cite{deVega:1977rk}. 
In Fig.~\ref{fig:cylinder}, 
we cut the torus along the $\beta$ cycle and arrange it as a straight tube along the $\alpha$ cycle, 
whose top and bottom are identified. 
Figures \ref{fig:cylinder} (a) and (b) 
show two configurations with the different lump charges $Q=1$ and $Q=2$, with the same twisting $P=1$ along the $\alpha$ cycle.
The preimage of ${\bf n} = (1,0,0)$ and ${\bf n} = (-1,0,0)$
are drawn as curves in Fig.~\ref{fig:cylinder} (a) and (b). 
We find in Fig.~\ref{fig:cylinder} (a) 
that the linking number of the preimages of ${\bf n}=(1,0,0)$ and 
${\bf n}=(-1,0,0)$ is one if we identify the bottom and the top. 
We thus have found that it carries the unit Hopf charge, {\it i.e.},
$C = P Q = 1$.

For the lump charge $Q=2$ in Fig.~\ref{fig:cylinder} (b), 
there exist two curves for the preimages of ${\bf n}=(1,0,0)$ at each slice $z=$ constant for $Q=2$,
each of which rotates clockwisely with the angle $\pi$ along the $\alpha$ cycle.
They are glued to each other to become one closed curve rotating $2\pi$ along the $\alpha$ cycle.
The same structure can be seen for the preimages of ${\bf n}=(-1,0,0)$ which is linked twice with the preimages of
${\bf n}=(1,0,0)$.
Consequently, the linking number of 
the two preimages is two and the Hopf charge for the lump charge $Q=2$ is two, {\it i.e.}, $C = P Q = 2$. 

The same is true for configuration with 
an arbitrary lump charge $Q$;
there exist $Q$ curves for  
the preimages of ${\bf n}=(\pm 1,0,0)$ at each slice.
They rotates clockwisely with the angle $2\pi/Q$ 
and are glued together to be one closed curve 
rotating $2\pi$ along the $\alpha$ cycle.
Then, the preimage of ${\bf n}=(1,0,0)$ 
has a linking number $Q$ with the preimage of 
${\bf n}=(-1,0,0)$ at the tube center, 
implying that the Hopf charge is $Q$ {\it i.e.}, $C = P Q = Q$. 
If we rotate the $U(1)$ modulus $P$ times along the $\alpha$ cycle,  
it further carries the Hopf charge $C=PQ$.

\begin{figure}
\begin{spacing}{0.8}
\vspace{-1.5cm}
\centering
\includegraphics[width=1\linewidth,keepaspectratio]{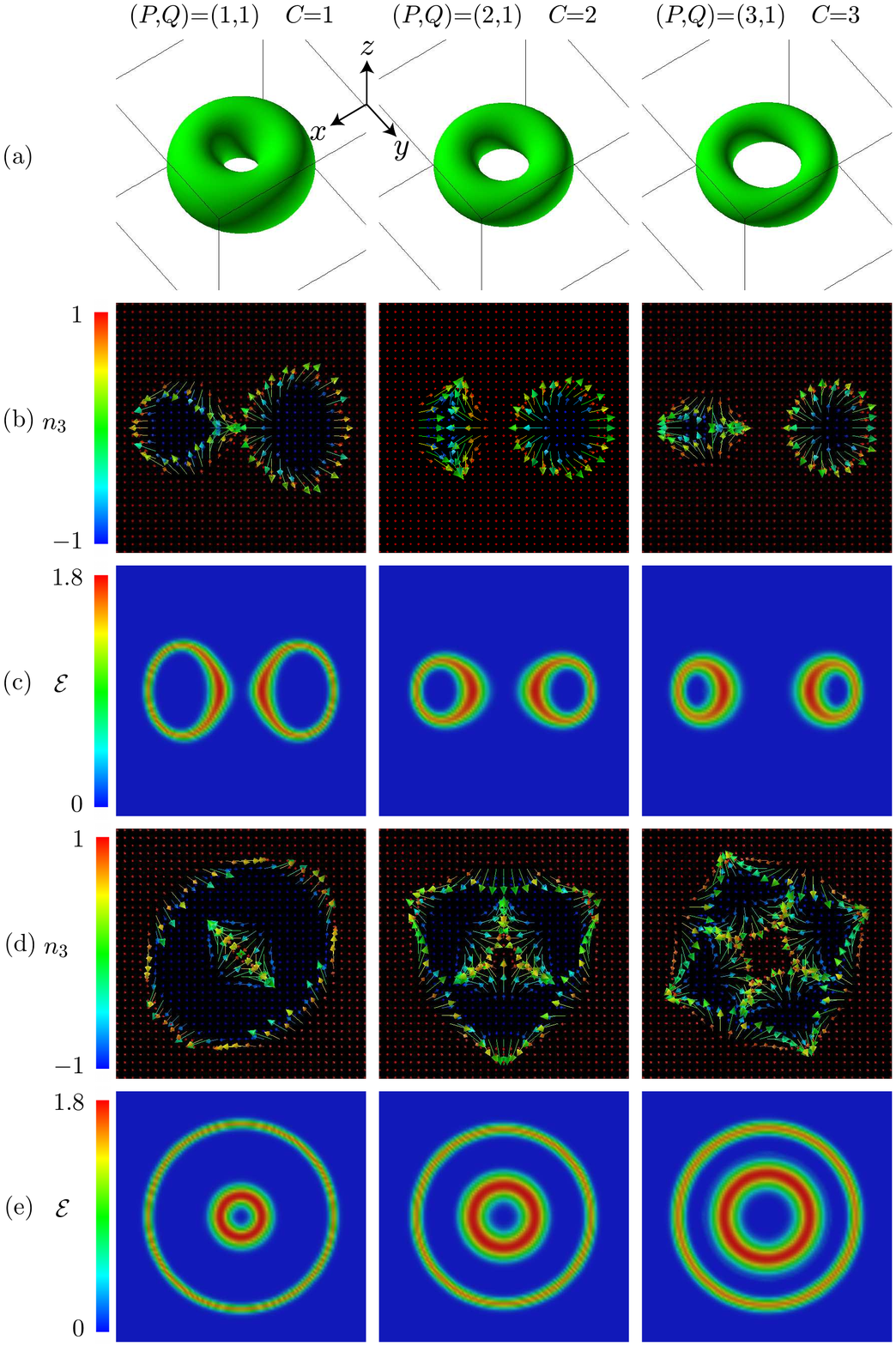}\\
\caption{
Toroidal domain walls with $(P,Q) = (1,1), (2,1)$ and $(3,1)$ 
for the left, middle and right figures, respectively.
(a): Isosurface plot of the $n_3 = 0$ region.
(b) The texture ${\bf n}(x)$ and 
(c) the total energy density $\mathcal{E}$ 
on the cross section of the $y$-$z$ plane.
(d) The texture ${\bf n}(x)$ and
(e) the total energy density $\mathcal{E}$ 
on the cross section of the $x$-$y$ plane.
The numerical box satisfies the Neumann boundary condition.
As numerical parameters, we fix $m^2 = 8000$, and $\kappa / L^4 = 0.2 \times 10^{-3}$.
\label{fig:wall-toroidal-q1}
}
\end{spacing}
\end{figure}
\begin{figure}
\begin{spacing}{0.8}
\vspace{-1cm}
\centering
\includegraphics[width=1\linewidth,keepaspectratio]{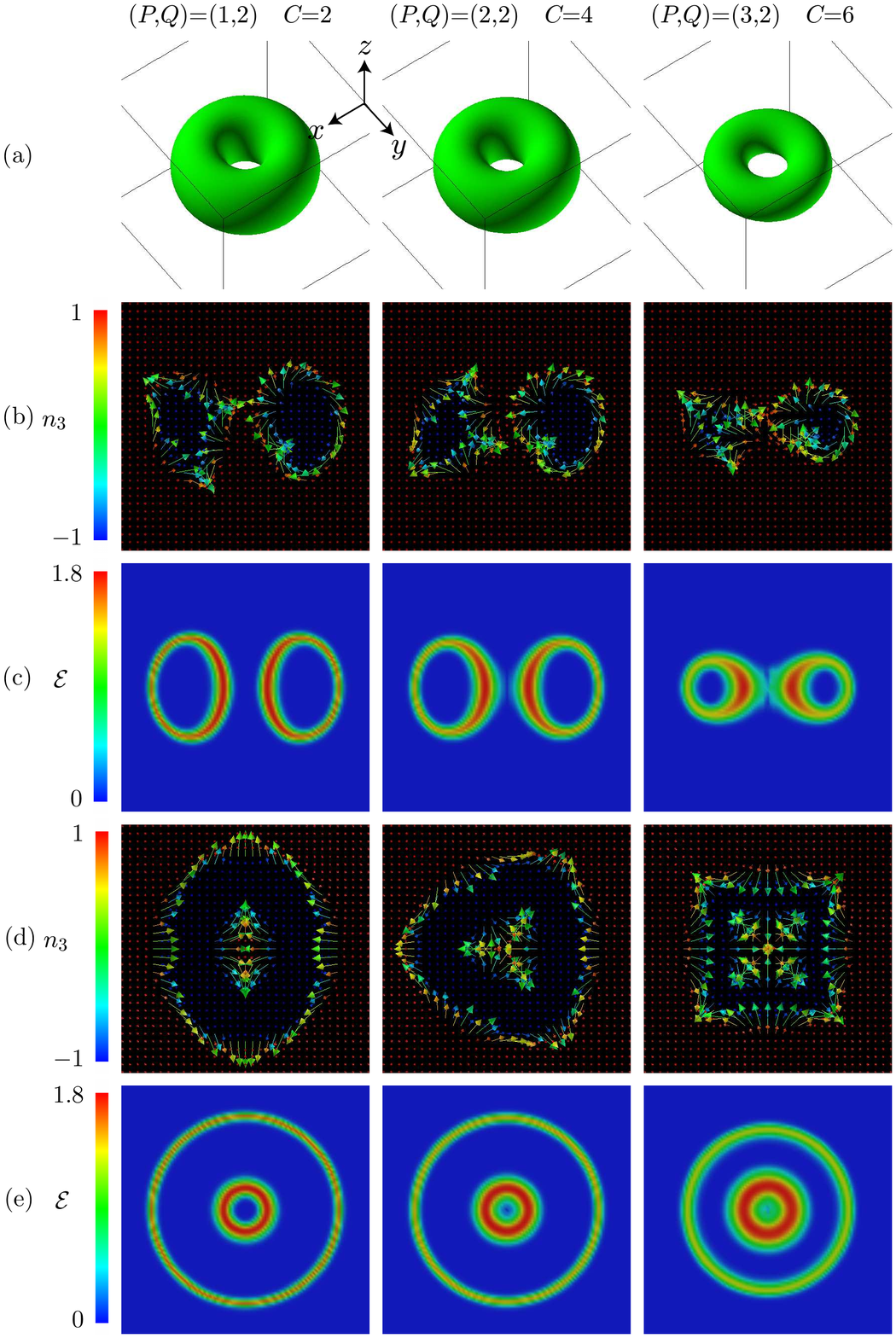}\\
\caption{
Toroidal domain walls with $(P,Q) = (1,2), (2,2)$ and $(3,2)$ 
for the left, middle and right figures, respectively.
We use $\kappa / L^4 = 0.1 \times 10^{-3}$, and the same $m^2$ and the boundary condition as those in Fig.~\ref{fig:wall-toroidal-q1}.
\label{fig:wall-toroidal-q2}
}
\end{spacing}
\end{figure}
\begin{figure}
\begin{spacing}{0.8}
\vspace{-1cm}
\centering
\includegraphics[width=1\linewidth,keepaspectratio]{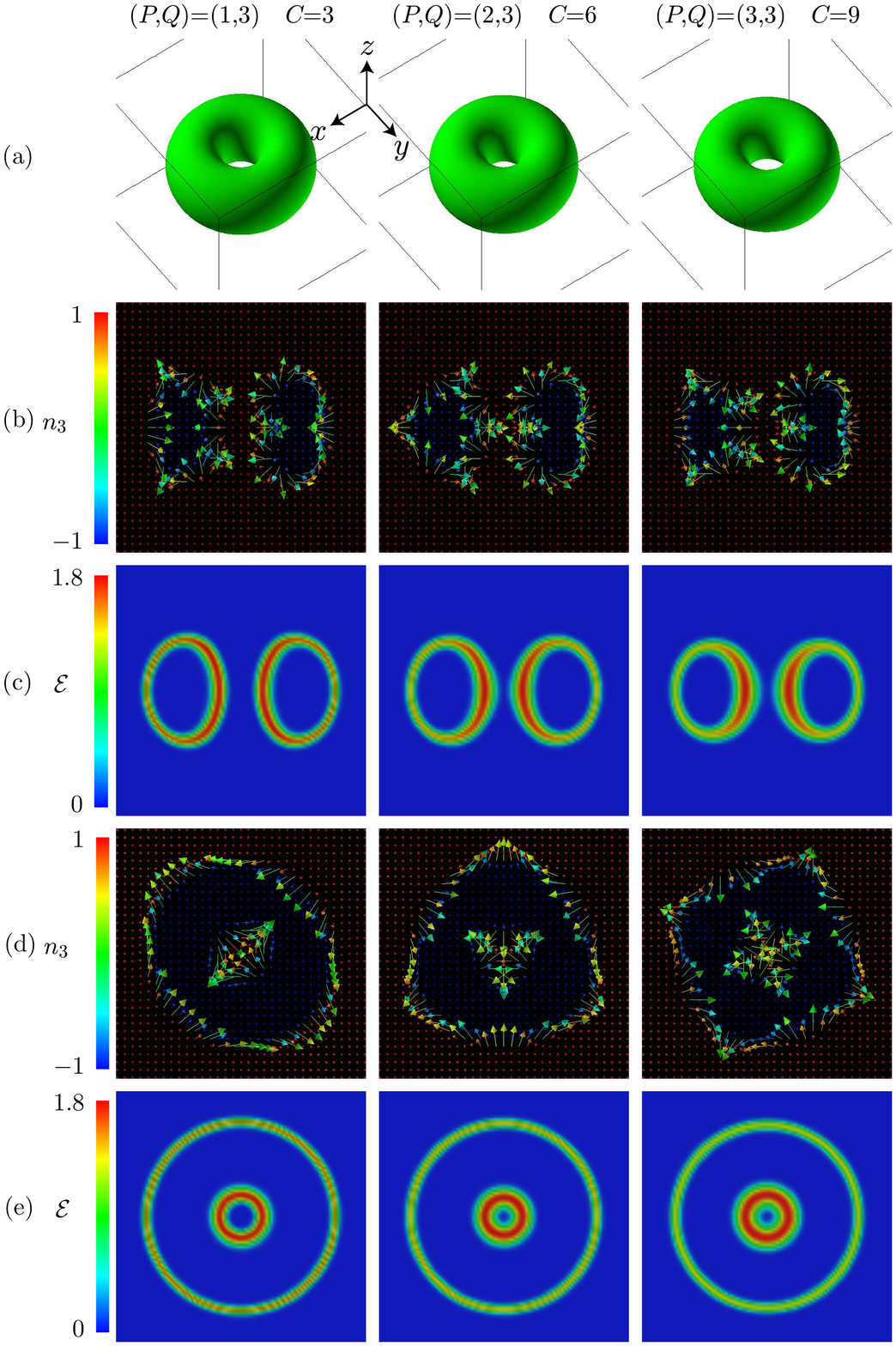}\\
\caption{
Toroidal domain walls with $(P,Q) = (1,3), (2,3)$ and $(3,3)$ 
for the left, middle and right figures, respectively.
We use $\kappa / L^4 = 0.5 \times 10^{-4}$, and the same $m^2$ and the boundary condition as those in Fig.~\ref{fig:wall-toroidal-q1}.
\label{fig:wall-toroidal-q3}
}
\end{spacing}
\end{figure}
In Fig.~\ref{fig:wall-toroidal-q1}, we present our numerical 
solutions of toroidal domain walls 
with $(P,Q) = (1,1), (2,1)$, and $(3,1)$.
The corresponding Hopf charges are 1, 2, 3.
One can see a pair of a lump (with the charge $Q$) and an anti-lump (with the charge $-Q$) 
in the section of the $y$-$z$ plane,  
as in Figs.~\ref{fig:wall-toroidal-q1} (b) and (c). 
On the other hand,  
 two wall rings with large and small radii 
with the common center are both lumps 
with the charge $P(>0)$ 
in the section of the $x$-$y$ plane,
as in  Fig.~\ref{fig:wall-toroidal-q1}(d) and (e).
For higher $P$, the size of the toroidal domain wall becomes slightly thinner as can be seen in the figure.

Figures \ref{fig:wall-toroidal-q2} and \ref{fig:wall-toroidal-q3} show our numerical solution of toroidal domain walls with higher $Q$, that is, $(P,Q) = (1,2), (2,2)$, and $(3,2)$ 
in Fig.~\ref{fig:wall-toroidal-q2} 
(the Hopf charges are $2,4,6$) and 
$(P,Q) = (1,3), (2,3)$, and $(3,3)$ in Fig.~\ref{fig:wall-toroidal-q3} (the Hopf charges are $3,6,9$). 
For higher $Q$, the toroidal domain wall becomes fatter because the underlying lump soliton becomes larger.
To make the sizes of toroidal domain walls within different $Q$ almost same, we use smaller $\kappa$ for higher $Q$ as well as the case of the wall rings.

Our toroidal domain walls classified 
by the two topological charges $(P,Q)$ are all stable. 
The Hopfions with the Hopf charge $C$ in our model are
further classified 
into a set of infinite series characterized by 
the two topological charges $(P,Q)$ with $C=PQ$.
On the other hand, conventional Hopfions in the original FS model are classified only by $C$; 
Configurations with different sets $(P,Q)$ with $C=PQ$ are topologically equivalent and can be deformed
to each other \cite{Hietarinta:2000ci}. 
In particular, configurations with the higher $Q$ 
would not be stable.
In our model, configurations with different 
sets $(P,Q)$ are topologically distinct and are all stable 
at the local minima of the energy.
Our numerical solutions imply 
the existence of a potential barrier among them. 
This is a new feature of our model 
which does not exist in the original FS model. 

\section{Discussion \label{sec:summary} }
If we give a linear time dependence of the $U(1)$ phase, 
the lumps on each section of the torus become Q-lumps \cite{Leese:1991hr} 
and our solution becomes a Q-torus 
which was studied without the Skyrme term in \cite{Bolognesi:2007zz}. 
In our case, the time dependence 
is not necessary for the stability, 
while it is necessary in their case because of 
the absence of the Skyrme term. 
The stability analysis of a Q-torus without the Skyrme term 
is an open question.

Our model can be promoted to a $U(1)$ gauge theory coupled 
with two massive charged complex scalar fields 
(with a four derivative term on the scalar fields, 
see Appendix B of \cite{Eto:2012qda}), 
in which lumps are promoted to semi-local vortices 
\cite{Vachaspati:1991dz}. 
In the strong gauge coupling limit, the gauge theory goes back 
to the massive $O(3)$ sigma model.
If we give a linear time dependence of the $U(1)$ modulus, 
it resembles a vorton \cite{Radu:2008pp}. 

Our model admits a D-brane soliton, namely, 
a domain wall on which lump-strings ending on from the both sides in the absence of the Skyrme term 
\cite{Gauntlett:2000de,Isozumi:2004vg}. 
This solution will be slightly modified by the presence of the Skyrme term. 
A configuration made of
a domain wall and an anti-domain wall 
stretched by lump-strings were considered 
in \cite{Nitta:2012kk}, in which it was discussed that 
such a configuration is unstable to decay,  
resulting in a creation of Hopfions studied in this paper.
Our model is the unique model admitting a domain wall, 
lump-strings and Hopfions as codimension one, two, and three solitons, respectively.  
As denoted above, the lump strings can end on 
the domain wall. 
It is also an interesting problem if a Hopfion can constitute a composite soliton with a domain wall or lump strings. 
Interaction between Hopfions and that between 
a Hopfion and a domain wall or a lump string 
also remain as future topics.

The FS model can be made supersymmetric 
if one adds another four derivative term 
\cite{Bergshoeff:1984wb,Eto:2012qda}.  
The potential term which we consider is well known 
in supersymmetric theory. 
However, the extra four derivative term tends to destabilize Hopfions at least in the absence of the potential term.

We can extend our model to the ${\bf C}P^N$ model or 
the Grassmannian sigma model.
With the appropriate (supersymmetric) potential term, 
the ${\bf C}P^N$ model (without the Skyrme term) is known to 
admit $N$ vacua and 
$N-1$ parallel domain walls interpolating them \cite{Gauntlett:2000ib}  
or junctions of these domain walls \cite{Eto:2005cp}.  
The Grassmannian sigma model 
with the (supersymmetric) potential term \cite{Arai:2003tc} 
admits non-Abelian domain walls \cite{Isozumi:2004jc} 
or non-Abelian wall junction \cite{Eto:2005fm}.
The ${\bf C}P^N$ model with the Skyrme-like four derivative terms was studied in \cite{Ferreira:2008nn}.  
The ${\bf C}P^N$ model with the Skyrme-like terms and the potential term may allow multi-layered toroidal domain walls, 
which remains as an open question.

\section*{Acknowledgements}

We thank the organizers of the conference 
``Quantized Flux in Tightly Knotted and Linked Systems," 
held in 3 - 7 December 2012 at Isaac Newton Institute 
for Mathematical Sciences, where this work was initiated. 
This work is supported in part by 
Grant-in-Aid for Scientific Research (Grant No. 22740219 (M.K.) and No. 23740198 and 25400268 (M.N.)) 
and the work of M. N. is also supported in part by 
the ``Topological Quantum Phenomena'' 
Grant-in-Aid for Scientific Research 
on Innovative Areas (No. 23103515 and 25103720)  
from the Ministry of Education, Culture, Sports, Science and Technology 
(MEXT) of Japan. 

\newcommand{\J}[4]{{\sl #1} {\bf #2} (#3) #4}
\newcommand{\andJ}[3]{{\bf #1} (#2) #3}
\newcommand{\AP}{Ann.\ Phys.\ (N.Y.)}
\newcommand{\MPL}{Mod.\ Phys.\ Lett.}
\newcommand{\NP}{Nucl.\ Phys.}
\newcommand{\PL}{Phys.\ Lett.}
\newcommand{\PR}{ Phys.\ Rev.}
\newcommand{\PRL}{Phys.\ Rev.\ Lett.}
\newcommand{\PTP}{Prog.\ Theor.\ Phys.}
\newcommand{\hep}[1]{{\tt hep-th/{#1}}}


\begin{thebibliography}{100}
\bibitem{Volovik2003}
G.~E.~Volovik,
{\it The Universe in a Helium Droplet}, 
Clarendon Press,  Oxford (2003).

\bibitem{Manton:2004tk}
  N.~S.~Manton and P.~Sutcliffe,
  ``Topological solitons,''
{\it  Cambridge, UK: Univ. Pr. (2004) 493 p}.

\bibitem{Vilenkin:2000}
A.~Vilenkin and E.~P.~S.~Shellard, 
{\it Cosmic Strings and Other Topological Defects}, (Cambridge Monographs on Mathematical Physics), Cambridge University Press (July 31, 2000).

\bibitem{Faddeev:1975}
L.~D.~Faddeev, Princeton preprint IAS-75-QS70.

\bibitem{Faddeev:1996zj} 
  L.~D.~Faddeev and A.~J.~Niemi,
  ``Knots and particles,''  Nature {\bf 387}, 58 (1997)  [hep-th/9610193].  

\bibitem{Nitta:2012kk} 
  M.~Nitta,
  ``Knots from wall--anti-wall annihilations with stretched strings,''  Phys.\ Rev.\ D {\bf 85}, 121701 (2012)  [arXiv:1205.2443 [hep-th]].  

\bibitem{Gladikowski:1996mb} 
  J.~Gladikowski and M.~Hellmund,
  ``Static solitons with nonzero Hopf number,''  Phys.\ Rev.\ D {\bf 56}, 5194 (1997)  [hep-th/9609035].  

\bibitem{Battye:1998pe} 
  R.~A.~Battye and P.~M.~Sutcliffe,
  ``Knots as stable soliton solutions in a three-dimensional classical field theory,''  Phys.\ Rev.\ Lett.\  {\bf 81}, 4798 (1998)  [hep-th/9808129];  
  ``Solitons, links and knots,''  Proc.\ Roy.\ Soc.\ Lond.\ A {\bf 455}, 4305 (1999)  [hep-th/9811077].  

\bibitem{Radu:2008pp} 
  E.~Radu and M.~S.~Volkov,
  ``Existence of stationary, non-radiating ring solitons in field theory: knots and vortons,''  Phys.\ Rept.\  {\bf 468}, 101 (2008)  [arXiv:0804.1357 [hep-th]].  

\bibitem{Hietarinta:2000ci} 
  J.~Hietarinta and P.~Salo,
  ``Ground state in the Faddeev-Skyrme model,''  Phys.\ Rev.\ D {\bf 62}, 081701 (2000).  

\bibitem{Sutcliffe:2007ui} 
  P.~Sutcliffe,
  ``Knots in the Skyrme-Faddeev model,''  Proc.\ Roy.\ Soc.\ Lond.\ A {\bf 463}, 3001 (2007)  [arXiv:0705.1468 [hep-th]].  

\bibitem{Foster:2010zb} 
  D.~Foster,
  ``Massive Hopfions,''  Phys.\ Rev.\ D {\bf 83}, 085026 (2011)  [arXiv:1012.2595 [hep-th]].  

\bibitem{Harland:2013uk} 
  D.~Harland, J.~Jaykka, Y.~Shnir and M.~Speight,
  ``Isospinning hopfions,''  arXiv:1301.2923 [hep-th].  

\bibitem{Battye:2013xf} 
  R.~A.~Battye and M.~Haberichter,
  ``Classically Isospinning Hopf Solitons,''  arXiv:1301.6803 [hep-th].  

\bibitem{deVega:1977rk} 
  H.~J.~de Vega,
  ``Closed Vortices and the HOPF Index in Classical Field Theory,''  
Phys.\ Rev.\ D {\bf 18}, 2945 (1978);  
  A.~Kundu and Y.~P.~Rybakov,
  ``Closed Vortex Type Solitons With Hopf Index,''  
J.\ Phys.\ A A {\bf 15}, 269 (1982).  

\bibitem{Abraham:1992vb} 
  E.~R.~C.~Abraham and P.~K.~Townsend,
  ``Q kinks,''  
Phys.\ Lett.\ B {\bf 291}, 85 (1992);  
  ``More on Q kinks: A (1+1)-dimensional analog of dyons,''  
Phys.\ Lett.\ B {\bf 295}, 225 (1992).  

\bibitem{Arai:2002xa} 
  M.~Arai, M.~Naganuma, M.~Nitta and N.~Sakai,
  ``Manifest supersymmetry for BPS walls in N=2 nonlinear sigma models,''  Nucl.\ Phys.\ B {\bf 652}, 35 (2003)  [hep-th/0211103];  
  ``BPS wall in N=2 SUSY nonlinear sigma model with Eguchi-Hanson manifold,''  In *Arai, A. (ed.) et al.: A garden of quanta* 299-325  [hep-th/0302028].  

\bibitem{Polyakov:1975yp} 
  A.~M.~Polyakov and A.~A.~Belavin,
  ``Metastable States of Two-Dimensional Isotropic Ferromagnets,''  
JETP Lett.\  {\bf 22}, 245 (1975)  [Pisma Zh.\ Eksp.\ Teor.\ Fiz.\  {\bf 22}, 503 (1975)].  

\bibitem{Nitta:2012kj} 
  M.~Nitta,
  ``Defect formation from defect--anti-defect annihilations,''  Phys.\ Rev.\ D {\bf 85}, 101702 (2012)  [arXiv:1205.2442 [hep-th]].  

\bibitem{Kobayashi:2013ju} 
  M.~Kobayashi and M.~Nitta,
  ``Jewels on a wall ring,''  Phys.\  Rev.\ D {\bf 87}, 085003 (2013)  [arXiv:1302.0989 [hep-th]].  


\bibitem{Piette:1994ug}
  B.~M.~A.~Piette, B.~J.~Schroers and W.~J.~Zakrzewski,
  ``Multi - Solitons In A Two-Dimensional Skyrme Model,''
  Z.\ Phys.\  C {\bf 65}, 165 (1995);
  ``Dynamics of baby skyrmions,''
  Nucl.\ Phys.\  B {\bf 439}, 205 (1995).

\bibitem{Weidig:1998ii}
  T.~Weidig,
  ``The baby Skyrme models and their multi-skyrmions,''
Nonlinearity {\bf 12}, 1489-1503 (1999). 

\bibitem{Bolognesi:2007zz} 
  S.~Bolognesi and M.~Shifman,
  ``Q torus in N=2 supersymmetric QED,''  Phys.\ Rev.\ D {\bf 76}, 125024 (2007)  [arXiv:0705.0379 [hep-th]].  

\bibitem{Nitta:2012wi} 
  M.~Nitta,
  ``Correspondence between Skyrmions in 2+1 and 3+1 Dimensions,''  Phys.\ Rev.\ D {\bf 87}, 025013 (2013)  [arXiv:1210.2233 [hep-th]];  
  M.~Nitta,
  ``Matryoshka Skyrmions,''  Nucl.\  Phys.\ B {\bf 872}, 62 (2013)  [arXiv:1211.4916 [hep-th]].  


\bibitem{Kudryavtsev:1997nw}
  A.~E.~Kudryavtsev, B.~M.~A.~Piette and W.~J.~Zakrzewski,
  ``Skyrmions and domain walls in (2+1) dimensions,''
  Nonlinearity {\bf 11}, 783 (1998);
  D.~Harland and R.~S.~Ward,
  ``Walls and chains of planar skyrmions,''
  Phys.\ Rev.\  D {\bf 77}, 045009 (2008).

\bibitem{Leese:1991hr} 
  R.~A.~Leese,
  ``Q lumps and their interactions,''  Nucl.\ Phys.\ B {\bf 366}, 283 (1991);  
  E.~Abraham,
  ``Nonlinear sigma models and their Q lump solutions,''  Phys.\ Lett.\ B {\bf 278}, 291 (1992).  

\bibitem{Eto:2012qda} 
  M.~Eto, T.~Fujimori, M.~Nitta, K.~Ohashi and N.~Sakai,
  ``Higher Derivative Corrections to Non-Abelian Vortex Effective Theory,''  Prog.\ Theor.\ Phys.\  {\bf 128}, 67 (2012)  [arXiv:1204.0773 [hep-th]].  

\bibitem{Vachaspati:1991dz}
  T.~Vachaspati and A.~Achucarro,
  ``Semilocal cosmic strings,''
  Phys.\ Rev.\  D {\bf 44}, 3067 (1991);
  A.~Achucarro and T.~Vachaspati,
  ``Semilocal and electroweak strings,''
  Phys.\ Rept.\  {\bf 327}, 347 (2000)
  [Phys.\ Rept.\  {\bf 327}, 427 (2000)]
  [arXiv:hep-ph/9904229].

\bibitem{Gauntlett:2000de} 
  J.~P.~Gauntlett, R.~Portugues, D.~Tong and P.~K.~Townsend,
  ``D-brane solitons in supersymmetric sigma models,''  
Phys.\ Rev.\ D {\bf 63}, 085002 (2001) 
  [hep-th/0008221];  
  M.~Shifman and A.~Yung,
  ``Domain walls and flux tubes in N=2 SQCD: D-brane prototypes,''  
Phys.\ Rev.\ D {\bf 67}, 125007 (2003)  [hep-th/0212293].  

\bibitem{Isozumi:2004vg} 
  Y.~Isozumi, M.~Nitta, K.~Ohashi and N.~Sakai,
  ``All exact solutions of a 1/4 Bogomol'nyi-Prasad-Sommerfield equation,''
  Phys.\ Rev.\ D {\bf 71}, 065018 (2005)
  [hep-th/0405129];
  M.~Eto, Y.~Isozumi, M.~Nitta, K.~Ohashi and N.~Sakai,
  ``Solitons in the Higgs phase: The Moduli matrix approach,''  J.\ Phys.\ A {\bf 39}, R315 (2006)  [hep-th/0602170];  
  M.~Eto, Y.~Isozumi, M.~Nitta and K.~Ohashi,
  ``1/2, 1/4 and 1/8 BPS equations in SUSY Yang-Mills-Higgs systems: Field theoretical brane configurations,''  Nucl.\ Phys.\ B {\bf 752}, 140 (2006)  [hep-th/0506257].  

\bibitem{Bergshoeff:1984wb} 
  E.~A.~Bergshoeff, R.~I.~Nepomechie and H.~J.~Schnitzer,
  ``Supersymmetric Skyrmions In Four-dimensions,''  Nucl.\ Phys.\ B {\bf 249}, 93 (1985);  
  L.~Freyhult,
  ``The supersymmetric extension of the Faddeev model,''
  Nucl.\ Phys.\  B {\bf 681}, 65 (2004)
  [arXiv:hep-th/0310261].


\bibitem{Gauntlett:2000ib} 
  J.~P.~Gauntlett, D.~Tong and P.~K.~Townsend,
  ``Multidomain walls in massive supersymmetric sigma models,''  Phys.\ Rev.\ D {\bf 64}, 025010 (2001)  [hep-th/0012178];  
  D.~Tong,
  ``The Moduli space of BPS domain walls,''  
Phys.\ Rev.\ D {\bf 66}, 025013 (2002)  [hep-th/0202012]. 

\bibitem{Eto:2005cp} 
  M.~Eto, Y.~Isozumi, M.~Nitta, K.~Ohashi and N.~Sakai,
  ``Webs of walls,''  Phys.\ Rev.\ D {\bf 72}, 085004 (2005)  [hep-th/0506135];   
  M.~Eto, Y.~Isozumi, M.~Nitta, K.~Ohashi, K.~Ohta and N.~Sakai,
  ``D-brane configurations for domain walls and their webs,''  AIP Conf.\ Proc.\  {\bf 805}, 354 (2006)  [hep-th/0509127];  
  M.~Eto, T.~Fujimori, T.~Nagashima, M.~Nitta, K.~Ohashi and N.~Sakai,
  ``Dynamics of Domain Wall Networks,''  Phys.\ Rev.\ D {\bf 76}, 125025 (2007)  [arXiv:0707.3267 [hep-th]].  

\bibitem{Arai:2003tc} 
  M.~Arai, M.~Nitta and N.~Sakai,
  ``Vacua of massive hyperKahler sigma models of nonAbelian quotient,''  Prog.\ Theor.\ Phys.\  {\bf 113}, 657 (2005)  [hep-th/0307274].  


\bibitem{Isozumi:2004jc} 
  Y.~Isozumi, M.~Nitta, K.~Ohashi and N.~Sakai,
  ``Construction of non-Abelian walls and their complete moduli space,''  
Phys.\ Rev.\ Lett.\  {\bf 93}, 161601 (2004)  [hep-th/0404198];  
  Y.~Isozumi, M.~Nitta, K.~Ohashi and N.~Sakai,
  ``Non-Abelian walls in supersymmetric gauge theories,''  
Phys.\ Rev.\ D {\bf 70}, 125014 (2004)  [hep-th/0405194];  
  M.~Eto, Y.~Isozumi, M.~Nitta, K.~Ohashi, K.~Ohta and N.~Sakai,
  ``D-brane construction for non-Abelian walls,''  
Phys.\ Rev.\ D {\bf 71}, 125006 (2005)  [hep-th/0412024];  
  M.~Eto, Y.~Isozumi, M.~Nitta, K.~Ohashi, K.~Ohta, N.~Sakai and Y.~Tachikawa,
  ``Global structure of moduli space for BPS walls,''  Phys.\ Rev.\ D {\bf 71}, 105009 (2005)  [hep-th/0503033].  

\bibitem{Eto:2005fm} 
  M.~Eto, Y.~Isozumi, M.~Nitta, K.~Ohashi and N.~Sakai,
  ``Non-Abelian webs of walls,''  Phys.\ Lett.\ B {\bf 632}, 384 (2006)  [hep-th/0508241]. 

\bibitem{Ferreira:2008nn}
  L.~A.~Ferreira,
  ``Exact vortex solutions in an extended Skyrme-Faddeev model,''
  JHEP {\bf 0905}, 001 (2009) [arXiv:0809.4303 [hep-th]];
  L.~A.~Ferreira, N.~Sawado, K.~Toda,
  ``Static Hopfions in the extended Skyrme-Faddeev model,''
  JHEP {\bf 0911}, 124 (2009)
  [arXiv:0908.3672 [hep-th]].
  L.~A.~Ferreira, P.~Klimas,
  ``Exact vortex solutions in a $CP^N$ Skyrme-Faddeev type model,''
  JHEP {\bf 1010}, 008 (2010)
  [arXiv:1007.1667 [hep-th]].

\end{thebibliography}
\end{document}